\documentclass[12pt]{article}
\usepackage[dvips]{graphicx}

\newcommand{\be}{\begin{equation}}
\newcommand{\ee}{\end{equation}}
\newcommand{\bea}{\begin{eqnarray}}
\newcommand{\eea}{\end{eqnarray}}

\begin{document}
\begin{titlepage}
 
 
\vspace{1in}
 
\begin{center}
\Large
{\bf Qualitative Analysis of Early Universe Cosmologies}
 
\vspace{1in}

\normalsize

\large{Andrew P. Billyard$^{1a}$, Alan A. Coley$^{1,2b}$
 \& James E. Lidsey$^{3c}$}

\normalsize
\vspace{.7in}

$^1${\em Department of Physics, \\ Dalhousie University, Halifax, NS, B3H 3J5,
Canada} \\

\vspace{.1in}

$^2${\em Department of Mathematics, Statistics and Computing Science, \\ 
Dalhousie University,
Halifax, NS, B3H 3J5,
Canada} \\

\vspace{.1in}

$^3${\em Astronomy Centre and Centre for Theoretical Physics, \\
University of Sussex, Brighton, BN1 9QJ, U. K.}

\end{center}
 
\vspace{.5in}
 
\baselineskip=24pt
\begin{abstract}
\noindent
A qualitative analysis is presented for a class of homogeneous 
cosmologies derived from the string effective action when a 
cosmological constant is present in the matter sector of the theory. 
Such a term has significant effects 
on the qualitative dynamics. 
For example, models exist which
undergo a series of oscillations between expanding and contracting 
phases due to the existence of a heteroclinic cycle in the phase space. 
Particular analytical solutions corresponding to 
the equilibrium points are also found. 

\end{abstract}

PACS NUMBERS: 98.80.Cq, 04.50.+h, 98.80.Hw
 
\vspace{.7in}
\noindent$^a$Electronic mail: jaf@mscs.dal.ca  \\
$^b$Electronic mail: aac@mscs.dal.ca \\
$^c$Electronic mail: jlidsey@astr.cpes.susx.ac.uk

\end{titlepage}
\section{Introduction}

Very early universe cosmology provides one of the few environments 
where the predictions of fundamental theories of physics, and
in particular string theories, can be 
investigated. String theory is the most promising 
candidate for a unified theory of the fundamental interactions. 
It introduces significant modifications to
the standard, hot big bang model based on conventional 
Einstein gravity and a study of string--inspired cosmologies 
is therefore important.

String theories predict 
the existence of a graviton, $g_{\mu\nu}$, 
a scalar `dilaton'  field, $\Phi$, and an 
antisymmetric two--form potential, $B_{\mu\nu}$, with a field 
strength $H_{\mu\nu\lambda} \equiv 
\partial_{[\mu} B_{\nu\lambda ]}$ \cite{eff,gsw}.
In four dimensions, the  three--form field strength is dual to a one--form, 
$\nabla_{\mu} \sigma$, such that   
$H^{\mu\nu\lambda} \equiv e^{\Phi} \epsilon^{\mu\nu\lambda\kappa}
\nabla_{\kappa} \sigma$, 
where $\epsilon^{\mu\nu\lambda\kappa}$ is the covariantly 
constant four--form \cite{sen}. 
The one--form may be interpreted as the gradient of 
a scalar `axion' field. The string field equations can 
then be derived from the effective action \cite{sen}
\begin{equation}
\label{sigmaaction}
S=\int d^4 x \sqrt{-g} e^{-\Phi} \left[ 
R +\left( \nabla \Phi  \right)^2  
-\frac{1}{2} 
e^{2 \Phi} \left( \nabla \sigma \right)^2 \right] +S_{\rm M}  ,
\end{equation}
where $S_{\rm M}$ represents the action for perfect fluid matter sources, 
$R$ is the Ricci curvature of the spacetime and   
$g\equiv {\rm det}g_{\mu\nu}$. The dilaton--graviton 
sector of action (\ref{sigmaaction}) 
may be interpreted as a Brans--Dicke theory, where the coupling 
parameter between the two fields takes the specific value $\omega =-1$
\cite{bransdicke}. 
The value of the dilaton field determines the effective value of Newton's 
`constant', $G_{\rm eff} \propto e^{\Phi}$.

The general solutions to the field 
equations of action (\ref{sigmaaction}) are known analytically 
when $S_{\rm M} =0$ for both the spatially flat and 
isotropic Friedmann--Robertson--Walker (FRW)
universes and the anisotropic Bianchi type I models 
\cite{clw,mv}. The purpose of the present paper is to qualitatively 
investigate the consequences of introducing a cosmological 
constant, $\Lambda_{\rm M}$, into the matter sector of Eq.  (\ref{sigmaaction}): 
\begin{equation}
\label{generalaction}
S=\int d^4 x \sqrt{-g} \left\{ e^{-\Phi} \left[ R +\left( \nabla \Phi
\right)^2 -\frac{1}{2} e^{2\Phi}
\left( \nabla \sigma \right)^2\right] - \Lambda_{\rm M} \right\}.
\end{equation}
This term may be interpreted as a 
perfect fluid matter stress with an equation 
of state $p=-\rho$. It could be generated
by a slowly moving scalar field, with a kinetic energy 
contribution dominated by a self--interaction potential, 
$p \approx -V \approx -\rho$. Analytical FRW solutions have not 
been found for this model when the axion 
field is trivial and $\Lambda_{\rm M} >0$ \cite{barrow,lidseyprd}. 
Moreover, the combined effects of the cosmological constant
and axion field have not been considered previously. 

We determine the general structure of the phase 
space of solutions for spatially flat FRW and axisymmetric 
Bianchi type I cosmologies 
derived from action (\ref{generalaction}) for arbitrary $\Lambda_{\rm M}$. 
This complements the work of Refs. 
\cite{gp,kmol,emw,kmo,BCL}, where the qualitative effects of introducing a 
cosmological constant, $\Lambda_{\rm M} \propto e^{-\Phi}$, 
into the gravitational sector of Eq. (\ref{sigmaaction}) were 
determined. 

The paper is organized as follows. In Section 2, the cosmological 
field equations and solutions for a zero cosmological constant 
are presented. The qualitative behaviour 
of the models with positive and negative $\Lambda_{\rm M}$ is determined in 
Sections 3 and 4, respectively. The phase portraits are 
interpreted in Section 5 and we conclude with a discussion in Section 6. 

\section{Cosmological Field Equations}

The metric for the Bianchi type I model may be written in the form 
\begin{equation}
\label{Imetric}
ds^2 =-dt^2 +h_{ab}dx^a dx^b , \qquad a,b =1,2,3,
\end{equation}
where $h_{ab}(t)$ is a function of cosmic time $t$ only and 
represents the metric on the surfaces of homogeneity. The axisymmetric 
model may be parametrized  by 
$h_{ab} =e^{2\alpha (t)} \left( e^{2\beta (t)} \right)_{ab}$, 
where $e^{3\alpha}$ denotes the effective spatial volume of the universe.  
The traceless, diagonal matrix $\beta_{ab} \equiv {\rm diag} 
[\beta, \beta, -2\beta ]$ determines the shear of the models
and we refer to $\beta$ as the shear parameter \cite{rs}. 
The spatially flat, isotropic FRW model is recovered in the 
limit where 
$\beta =0$ and, in this case, $e^{\alpha}$ represents
the scale factor of the universe. 

Substituting the metric (\ref{Imetric}) into the 
action (\ref{generalaction}) and integrating over the spatial variables 
implies that 
\begin{equation}
\label{taction}
S=\int dt ~~e^{3\alpha} \left\{ e^{-\Phi} \left[ 6\dot{\alpha}\dot{\Phi} 
-6\dot{\alpha}^2 +6\dot{\beta}^2 -\dot{\Phi}^2 +
\frac{1}{2} e^{2\Phi} \dot{\sigma}^2 \right] -\Lambda_{\rm M} \right\}   ,
\end{equation}
where the co-moving volume has been normalized to unity without 
loss of generality and a dot denotes differentiation with respect 
to $t$.  
The field equations derived from Eq. (\ref{taction}) are given by  
\begin{eqnarray}
\label{r1}
\ddot{\alpha}  &=&\dot{\alpha} \dot{\varphi} +\dot{\varphi}^2 -3\dot{\alpha}^2 
-6 \dot{\beta}^2 -\frac{3}{2} \Lambda_{\rm M}  e^{\varphi +3\alpha} \\
\label{r2}
\ddot{\varphi} &=&3\dot{\alpha}^2  + 6 \dot{\beta}^2 +\frac{1}{2} 
\Lambda_{\rm M}  
e^{\varphi +3\alpha}  \\
\label{r3}
\ddot{\sigma} &=&-(\dot{\varphi}  +6 \dot{\alpha}  ) \dot{\sigma} \\
\label{r4}
\ddot{\beta} &=&\dot{\beta} \dot{\varphi}, 
\end{eqnarray} 
where 
\begin{equation}
\label{shifted}
\varphi \equiv \Phi -3\alpha
\end{equation}
defines the  `shifted' dilaton field
and the generalized  Friedmann constraint takes the form 
\begin{equation}
\label{frr}
3\dot{\alpha}^2 -\dot{\varphi}^2 +6 \dot{\beta}^2  +\frac{1}{2} \dot{\sigma}^2 
e^{2\varphi +6 \alpha} + 
\Lambda_{\rm M}  e^{\varphi +3\alpha} =0   .
\end{equation} 

Eqs. (\ref{r1})--(\ref{frr}) may be simplified by introducing the new 
time coordinate
\begin{equation}
\frac{d}{d \theta} \equiv e^{-(\varphi +3\alpha )/2} \frac{d}{dt} 
\end{equation}
and employing the generalized Friedmann constraint equation (\ref{frr}) 
to eliminate the axion field. The remaining  
field equations are then given by 
\begin{eqnarray}
\label{r5}
\alpha'' &=& \varphi'^2 -\frac{9}{2} \alpha'^2 +\frac{1}{2} \alpha'
\varphi' 
-6 \beta'^2 -\frac{3}{2} \Lambda_{\rm M}  \\
\label{r6}
\varphi'' &=&3\alpha'^2 +6\beta'^2 -\frac{1}{2} \varphi'^2 -
\frac{3}{2} \alpha' \varphi' +\frac{1}{2} \Lambda_{\rm M}   \\
\label{r7}
\beta'' &=& \frac{1}{2} \beta' \left( 
\varphi' -3 \alpha'  \right) ,
\end{eqnarray} 
where a prime denotes differentiation with 
respect to $\theta$. 

The general solution to Eqs. (\ref{r1})--(\ref{frr}) 
is known when the cosmological constant vanishes 
\cite{clw}. It is given by 
\begin{eqnarray}
e^{\alpha} &=& e^{\alpha_*} \left| \frac{s}{s_*} \right|^{1/2}
\left[ \left| \frac{s}{s_*} \right|^r +\left| 
\frac{s}{s_*} \right|^{-r} \right]^{1/2} \nonumber \\
e^{\Phi} &=&\frac{e^{\Phi_*}}{2} \left[ 
\left| \frac{s}{s_*} \right|^r + 
\left| \frac{s}{s_*} \right|^{-r} \right] \nonumber \\
\sigma &=& \sigma_* \pm e^{-\Phi_*} 
\left[ \frac{ |s/s_*|^{-r} -|s/s_*|^r}{|s/s_*|^{-r} + 
|s/s_*|^r} \right] \nonumber \\
\label{dma}
e^{\beta} &=&e^{\beta_*} \left| \frac{s}{s_*} \right|^q   ,
\end{eqnarray}
where $s\equiv \int^t dt' e^{-\alpha (t')}$ is 
conformal time, $\{ \alpha_* , s_* , \Phi_* , \sigma_* , \beta_* \}$ 
are arbitrary constants and $\{ r, q \}$ satisfy the 
constraint equation $ r = (3-12q^2)^{1/2}$. 

The solutions to Eqs. (\ref{r1})--(\ref{frr}) 
for a trivial axion field and zero cosmological constant have a power-law form:
\begin{eqnarray}
\label{dmvcosmic}
e^{\alpha}  & = & e^{\alpha_*} \left| t \right|^{\pm h_*} \nonumber \\
e^\Phi & = &e^{\Phi_*}\left|t\right|^{\pm3h_*-1} \nonumber \\
e^\beta & =& e^{\beta_*}\left|t\right|^{\pm\sqrt{(1-3h_*^2)/6}} ,
\end{eqnarray}
where $h_*$ is a constant such that $|h_*| \le 
1\sqrt{3}$. The solution (\ref{dma}) 
asymptotes to these power-law models at early and late times and 
the axion field is therefore 
dynamically negligible in these limits. When an axion field it present, 
as in Eq. (\ref{dma}),  
the universe undergoes a smooth transition between 
the two power-law solutions 
(\ref{dmvcosmic}) and exhibits a bounce when $s \approx s_*$. 
In the isotropic limit, $h_*^2 =1/3$, and the time--reversal of the 
$e^{\alpha} \propto |t|^{-1/\sqrt{3}}$ solution is inflationary. It  
corresponds to the pre--big bang cosmology, where  
the inflationary expansion is driven by the kinetic energy 
of the dilaton field \cite{pbb}. 

In the next section we determine the phase portraits 
for the generalized model with a non--trivial axion field 
and $\Lambda_{\rm M} >0$. The effect of the cosmological constant on 
the solutions (\ref{dma}) can then 
be established.  

\section{Positive Cosmological Constant}

When $\Lambda_{\rm M} > 0$, 
we can rewrite Eqs. (\ref{r5})--(\ref{r7}) using new variables 
defined by
\begin{equation}
\label{4.10}
h \equiv \alpha',  \qquad \psi \equiv \varphi' , \qquad 
N \equiv \beta'  .
\end{equation}
Eq. (\ref{frr}) then implies that  
\begin{equation}
\psi^2 \ge 3h^2 +6 N^2 +\Lambda_{\rm M}  \ge 0
\end{equation} 
and consequently we may normalize with $\psi$. We therefore 
define
\begin{eqnarray}
\label{x}
x &\equiv& \frac{\sqrt{3} h}{\psi} \\
\label{y}
y &\equiv& \frac{6 N^2}{\psi^2} \\
\label{z}
z &\equiv& \frac{\Lambda_{\rm M}}{\psi^2} \\
\label{theta}
\frac{d}{d\Theta} &\equiv& \frac{1}{\psi} \frac{d}{d\theta} 
\end{eqnarray}
and assume that $\psi>0$. (The case 
$\psi<0$ is related to a time-reversal of the system and the qualitative 
behaviour is similar).
The three--dimensional system (\ref{r5})--(\ref{r7}) is 
therefore given by 
\begin{eqnarray}
\frac{dx}{d\Theta} &= & (x+\sqrt{3})[1-x^2-y-z] +\frac{1}{2} z[x-\sqrt{3}] 
\label{xp} \\
\frac{dy}{d\Theta} &= & 2y\left\{[1-x^2-y-z]+\frac{1}{2}z\right\} 
\label{yp} \\
\frac{dz}{d\Theta} &= & 2z\left\{[1-x^2-y-z] 
-\frac{1}{2}(1-z-\sqrt{3}x)\right\} \label{zp}   .
\end{eqnarray}

It follows from the definitions (\ref{x})--(\ref{z}) 
that the phase space is
bounded with $0\leq \{ x^2,y,z\}\leq 1$ subject to 
the constraint $1-x^2-y-z\geq0$.  The
invariant set $1-x^2-y-z=0$ corresponds to a zero axion field.  The
dynamics of the system (\ref{xp})--(\ref{zp}) is determined primarily
by the dynamics in the invariant sets $y=0$ and $z=0$. 
These correspond
to a zero shear parameter and a zero cosmological constant, respectively. 
The dynamics is also determined by
the fact that the right-hand side of Eq. (\ref{yp}) is positive--definite
so that $y$ is a monotonically increasing function. This 
guarantees that there are no closed or
recurrent orbits in the three-dimensional phase space. 

\subsection{Isotropic Model for $\Lambda_{\rm M}>0$}

The isotropic FRW cosmology 
corresponds to the invariant set $y=0$, where the shear parameter is trivial.  
The system (\ref{xp})-(\ref{zp}) reduces to the following plane system
in this case:
\begin{eqnarray}
\label{dxdTh}
\frac{dx}{d\Theta} &= & (x+\sqrt{3})[1-x^2-z] +\frac{1}{2} z[x-\sqrt{3}] \\
\label{dzdTh}
\frac{dz}{d\Theta} &= & 2z\left\{[1-x^2-z] 
-\frac{1}{2}(1-z-\sqrt{3}x)\right\} .
\end{eqnarray}
The equilibrium points and their associated eigenvalues are given by
\begin{eqnarray}
S_1: & x=-1, z=0; & \lambda_1=2(\sqrt{3}-1), \quad \lambda_2=-(1+\sqrt{3}) \\
S_2: & x= 1, z=0; & \lambda_1=-2(\sqrt{3}+1), \quad \lambda_2= (\sqrt{3}-1) \\
F: & x=-\frac{1}{3\sqrt{3}}, z=\frac{16}{27}; & \lambda_{1,2}= \frac{1}{3} \pm 
\frac{i}{9}\sqrt{231}.
\end{eqnarray}
The points $S_1$ and $S_2$ are saddles and $F$ is a repelling focus. 
The phase portrait is given in Fig. 1.
 \begin{figure}[htp]
  \centering
   \includegraphics*[width=3in]{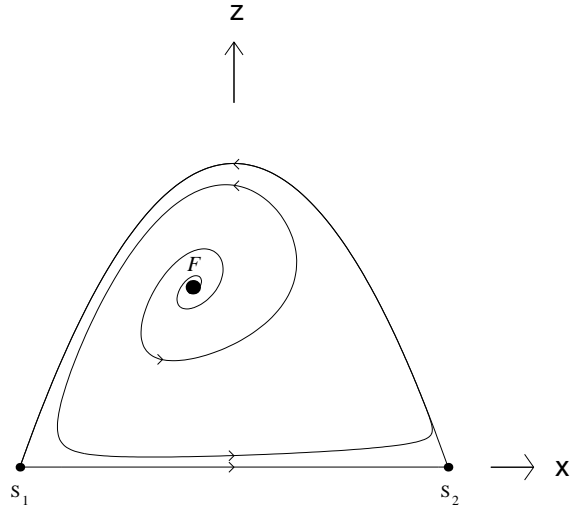}
\caption{{\em\noindent Phase portrait of the system
(\ref{dxdTh})--(\ref{dzdTh}), corresponding to the isotropic 
FRW model with $\Lambda_{\rm M} >0$.
Equilibrium points are denoted
by dots and the labels in all figures correspond to those equilibrium
points (and hence the exact solutions they represent) discussed in the
text.  We shall adopt the convention throughout that large black dots
represent sources (i.e., repellors), large grey-filled dots represent
sinks (i.e., attractors), and small black dots represent saddles.  Note that in
this phase space orbits are future asymptotic to a heteroclinic cycle. }} 
 \end{figure}

In the invariant set $1-x^2-z=0$, corresponding to the case of a
zero axion field, Eqs. (\ref{dxdTh}) and (\ref{dzdTh}) reduce to the
single ordinary differential equation: 
\begin{equation}
\label{invariantset}
\frac{dx}{d\Theta} = \frac{1}{2}\left(1-x^2\right)\left(x-\sqrt 3\right),
\end{equation}
which can be integrated to yield an exact solution in terms of 
$\Theta$--time.

\subsection{Anisotropic Model for $\Lambda_{\rm M}>0$}

In the full system (\ref{xp})-(\ref{zp}), 
corresponding to the anisotropic model with a non--trivial 
shear parameter, there exists the 
isolated equilibrium point (and their associated eigenvalues)
\begin{eqnarray}
\nonumber
F: && x=-\frac{1}{3\sqrt{3}},\qquad y=0, \qquad z=\frac{16}{27} \\
   && \left(\lambda_1,\lambda_1, \lambda_3\right)=\left(\frac{1}{3} +
	\frac{i}{9}\sqrt{231}, \frac{1}{3} - \frac{i}{9}\sqrt{231}, \frac{4}{3}
	\right) \label{F_point}
\\ \nonumber
W: && y=1-x^2, \qquad z=0 \qquad \mbox{($x$ arbitrary)} \\
   && \left(\lambda_1,\lambda_1, \lambda_3\right)=\left( -2\sqrt 
	3\left[x+\frac{1}{\sqrt{3}}\right], 
	\sqrt 3\left[x-\frac{1}{\sqrt{3}} \right], 0\right)\label{W_point}
\end{eqnarray}
Hence, $F$ is a global source.  The set $W$ lies in the invariant set
$z=0$ on the boundary $y=1-x^2$.  Points on $W$ with $x\in
(-1/\sqrt{3},1/\sqrt{3})$ are local sinks, while the remaining points
are saddles in the full three-dimensional phase space.  (In the
invariant set $z=0$ equilibrium points with $x\in [-1,-1/\sqrt{3})$
are repelling and those with $x\in(-1/\sqrt{3},1]$ are attracting).
The phase portrait for this system is given in Fig. 2 and table 
\ref{points1_asymp} lists each equilibrium set and its stability..
 \begin{figure}[htp]
  \centering
   \includegraphics*[width=5in]{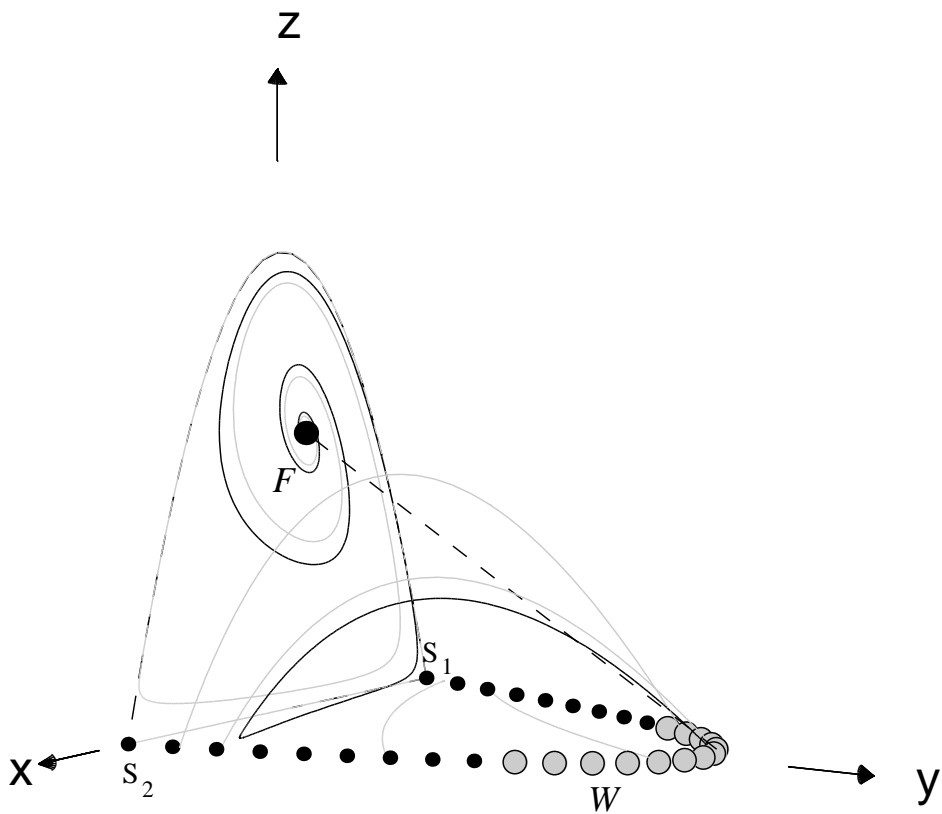}
\caption{{\em Phase portrait of the system (\ref{xp})--(\ref{zp})
corresponding to the axisymmetric Bianchi type I cosmology with
$\Lambda_{\rm M} >0$.  Note that $W$ denotes a line of non--isolated
equilibrium points.  The dashed line represents the exact
$x=\mbox{constant}$ solution (\ref{xfixed})--(\ref{zprime}).  See
caption to Fig. 1 for notation.  Arrows on the trajectories in this
three--dimensional phase space have been suppressed since the
direction of increasing time is clear using this notation.  Grey lines
represent typical trajectories found within the two-dimensional
invariant sets, dashed black lines are those trajectories along the
intersection of the invariant sets, and solid black lines are typical
trajectories within the full three-dimensional phase space.  }}
 \end{figure}

\begin{table}[hbt]
\begin{center}
\begin{tabular}{|l|l|}\hline
Equilibrium Point& Stability \\
\hline \hline
$F$: \ \  {\small Eq. (\ref{F_point})} & Repellor (Source) \\
\hline
$W$: \ {\small Eq. (\ref{W_point})} & Attractor (Sink) for 
	$\frac{-1}{\sqrt{3}}<x<\frac{1}{\sqrt{3}}$ \\
	& Saddle otherwise
\\ \hline
\end{tabular}
\end{center}
\caption{{\em Equilibrium sets for anisotropic model with
$\Lambda_{\rm M}>0$, and their stability (the equations where each set
is defined is also listed).}\label{points1_asymp}}
\end{table}

\

We note that there exists an exact, 
anisotropic solution of 
Eqs. (\ref{xp})--(\ref{zp}), where 
\begin{equation}
x=-\frac{1}{3\sqrt 3}= \mbox{constant} \label{xfixed}
\end{equation}
and 
\begin{equation}
y=-\frac{13}{8}\left(z-\frac{16}{27}\right)  . \label{yline}
\end{equation}
This implies that 
\begin{equation}
\label{intex}
\frac{dz}{d\Theta} = \frac{9}{4}z\left(z-\frac{16}{27}\right) \label{zprime}
\end{equation}
and Eq. (\ref{intex}) 
can be integrated explicitly in terms of $\Theta$--time.

In the following Section we determine the 
effects of a negative cosmological constant. This 
allows a direct comparison to be made with the 
$\Lambda_{\rm M} >0$ models considered above. 

\section{Negative Cosmological Constant}

In the case where $\Lambda_{\rm M} <0$, 
the generalized Friedmann constraint equation (\ref{frr}) implies that 
\begin{equation}
\psi^2-\Lambda_{\rm M}  \geq 3h^2 +6N^2 \geq 0 .
\end{equation}
We may therefore normalize by employing the quantity 
$\sqrt{\psi^2-\Lambda_{\rm M}}$. Defining the new variables
\begin{eqnarray}
\label{theu}
u&\equiv&\frac{\sqrt{3}h}{\sqrt{\psi^2-\Lambda_{\rm M}}} \\
\label{thev}
v&\equiv&\frac{\psi}{\sqrt{\psi^2-\Lambda_{\rm M}}} \\
\label{thew}
w&\equiv&\frac{6N^2}{\psi^2-\Lambda_{\rm M}},
\end{eqnarray}
where $0\leq\{u^2,v^2,w\}\leq1$, and the new time variable 
\begin{equation}
\frac{d}{d\Xi}=\frac{1}{\sqrt{\psi^2-\Lambda_{\rm M}}}\frac{d}{d\theta}
\end{equation}
implies that Eqs. (\ref{r5})-(\ref{r7}) become
\begin{eqnarray}
\frac{du}{d\Xi} & = & \frac{\sqrt 3}{2}\left(1-u^2\right)\left(1-v^2\right) 
	+ \left(1-u^2-w\right)\left(\sqrt 3 +uv\right) \label{dudXi} \\
\frac{dv}{d\Xi} & = & -\frac{1}{2}\left(1-v^2\right)\left(1-2u^2-2w
	+\sqrt{3}uv\right)    \label{dvdXi}\\
\frac{dw}{d\Xi} & = & w\left[ 2v\left(1-u^2-w\right)-\sqrt 3
	u\left(1-v^2\right)\right]. \label{dwdXi}
\end{eqnarray}
The phase space is bounded by the sets $v=\pm1$ and $w=1-u^2$, where the 
latter corresponds to a zero axion field.  The dynamics is
determined by the fact that the right-hand side of Eq. (\ref{dudXi}) is
positive definite so that $u$ is a monotonically increasing function.

\subsection{Isotropic Model for $\Lambda_{\rm M} <0$}

In the invariant set $w=0$, corresponding to the isotropic FRW model
$(\beta =0)$, the
system (\ref{dudXi})--(\ref{dwdXi}) reduces to the following
two-dimensional system:
\begin{eqnarray}
\frac{du}{d\Xi} & = & 
\frac{1}{2}\left(1-u^2\right)\left(2uv+\sqrt{3}[3-v^2]\right)
    \label{du} \\
\frac{dv}{d\Xi} & = & 
\frac{1}{2}\left(1-v^2\right)\left(2u^2-1-\sqrt{3}uv\right).
    \label{dv*}
\end{eqnarray}

The lines $u^2=1$ and $v^2=1$ are invariant sets, containing four equilibrium
points $S_{u,v}$. These points are all saddles 
and are  located 
at the intersections of the lines. Their eigenvalues are given by 
\begin{eqnarray}
S_{ 1, 1}: &  & \lambda_1=\sqrt3 -1, \quad \lambda_2=-2(\sqrt3+1) \\
S_{-1, 1}: &  & \lambda_1=2(\sqrt3 -1), \quad \lambda_2=-(\sqrt3+1) \\
S_{ 1,-1}: &  & \lambda_1=-2(\sqrt3-1), \quad \lambda_2= (\sqrt3+1)\\
S_{-1,-1}: &  & \lambda_1=-(\sqrt3-1), \quad \lambda_2=2(\sqrt3+1).
\end{eqnarray}
The remaining two equilibrium points and their eigenvalues are 
\begin{eqnarray}
R:(u_-,v_-)=\left(-1, -\frac{1}{\sqrt{3}}\right) & &  \lambda_1=
           \frac{1}{\sqrt{3}}, \quad \lambda_2= \frac{10}{\sqrt3}  \\
A:(u_+,v_+)=\left( 1,  \frac{1}{\sqrt{3}}\right)  & & \lambda_1=
          -\frac{1}{\sqrt{3}}, \quad \lambda_2=-\frac{10}{\sqrt3}.  
\end{eqnarray}          
Consequently, $R$ is a source and $A$ is a sink.  
Fig. 3 depicts the phase plane of the system (\ref{du})--(\ref{dv*}). 
 \begin{figure}[htp]
  \centering
   \includegraphics*[width=3in]{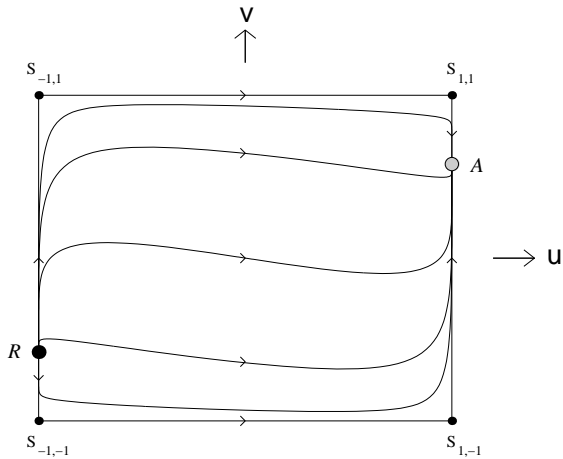}
\caption{{\em  Phase portrait of the system
(\ref{du})--(\ref{dv*}) corresponding to the isotropic FRW 
model with $\Lambda_{\rm M} <0$.  See caption to Fig. 1.}}
 \end{figure}

\subsection{Anisotropic Model for $\Lambda_{\rm M}<0$}

In the full system (\ref{dudXi})--(\ref{dwdXi}) with a non-trivial
shear parameter, the equilibrium points and their respective eigenvalues
are:
\begin{eqnarray}
\nonumber
W^\pm : && v=\pm 1, u^2+w=1; \\ 
        && (\lambda_1,\lambda_2,\lambda_3) = \left( 0,
        \sqrt{3}\left[u\mp\frac{1}
        {\sqrt3}\right], -2\sqrt 3\left[u\pm\frac{1}{\sqrt3}
        \right]\right) \label{Wpm_point}\\
\nonumber
R: && v=-\frac{1}{\sqrt3}, u=-1, w=0; \\
   && (\lambda_1,\lambda_2,\lambda_3) = \frac{1}{\sqrt3}
        \left(1, 2, 10\right) \label{R_point}\\
\nonumber
A: && v=\frac{1}{\sqrt3}, u=1, w=0; \\
   && (\lambda_1,\lambda_2,\lambda_3) = -\frac{1}{\sqrt3}
        \left(1, 2, 10\right). \label{A_point}
\end{eqnarray}
The saddle points $S_{\pm 1,-1}$ in subsection 4.1 are the
endpoints to the line $W^-$. This 
line represents early--time attracting solutions for
$-1/\sqrt3<u<1/\sqrt3$ and saddles otherwise. The saddle points
$S_{\pm 1,1}$ are the endpoints to the line $W^+$. This corresponds 
to late--time attracting
solutions for $-1/\sqrt3<u<1/\sqrt3$ and saddles otherwise.  
Hence, there are 
two early--time attractors given by the point $R$ and the line $W^-$
for $-1/\sqrt3<u<1/\sqrt3$. There are also 
two late--time attractors corresponding to the point $A$
and the line $W^+$ for $-1/\sqrt3< u <1/\sqrt3$. Fig. 4 depicts the
three--dimensional phase space and table \ref{points2_asymp} lists each 
equilibrium set and its stability.
 \begin{figure}[htp]
  \centering
   \includegraphics*[width=5in]{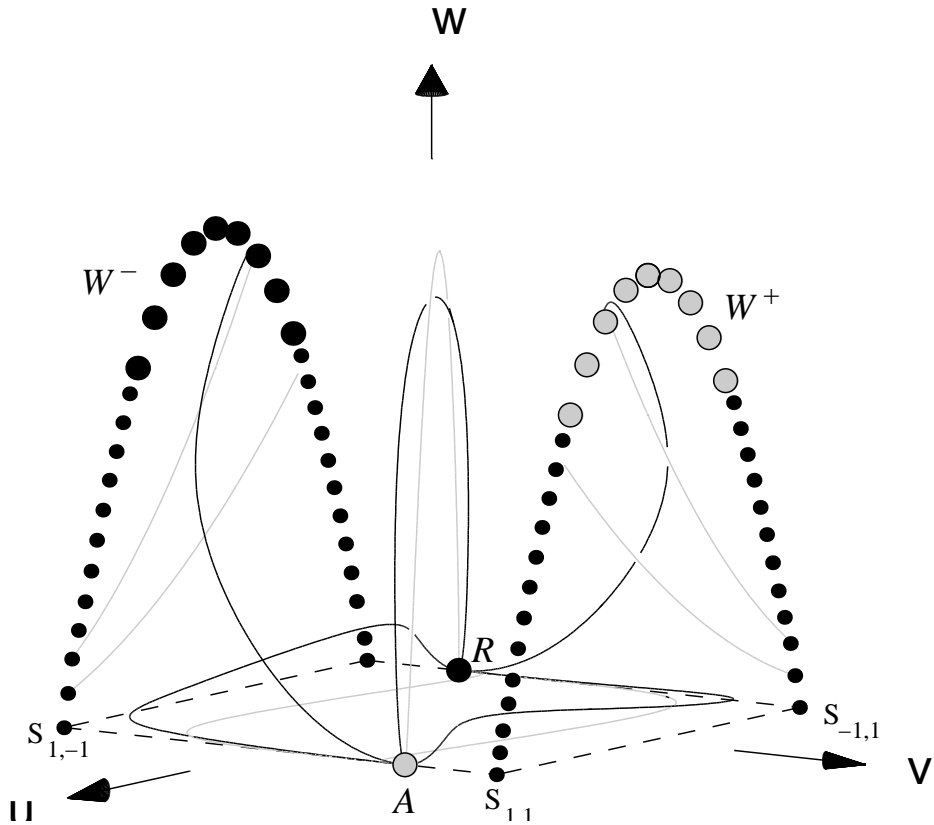}
\caption{{\em Phase portrait of the system 
(\ref{dudXi})--(\ref{dwdXi})
corresponding to the axisymmetric Bianchi type I 
model with $\Lambda_{\rm M} <0$.  Note that $W^\pm$ denote lines of
non-isolated equilibrium points.  See captions to Figs. 1 and 2.}}
 \end{figure}

\begin{table}[hbt]
\begin{center}
\begin{tabular}{|l|l|}\hline
Equilibrium Point& Stability \\
\hline \hline
$W^-$: \ {\small Eq. (\ref{Wpm_point})} & Repellor (Source) for 
	$\frac{-1}{\sqrt{3}}<u<\frac{1}{\sqrt{3}}$ \\
	& Saddle otherwise
\\ \hline
$W^+$: \ {\small Eq. (\ref{Wpm_point})} & Attractor (Sink) for 
	$\frac{-1}{\sqrt{3}}<u<\frac{1}{\sqrt{3}}$ \\
	& Saddle otherwise
\\ \hline
$R$: \ \ \ {\small Eq. (\ref{R_point})} & Repellor (Source)  
\\ \hline
$A$: \ \ \ {\small Eq. (\ref{A_point})} & Attractor (Sink)  
\\ \hline
\end{tabular}
\end{center}
\caption{{\em Equilibrium sets for anisotropic model with
$\Lambda_{\rm M}<0$, and their stability (the equations where each set
is defined is also listed).}\label{points2_asymp}}
\end{table}

\

This concludes the derivation
of the phase portraits for the spatially flat and 
homogeneous cosmologies derived from Eq. (\ref{generalaction}). 
We proceed in the following Section to discuss their properties. 

\section{Interpretation of the Phase Portraits}

The dynamics of the isotropic cosmology described by the system
(\ref{dxdTh})--(\ref{dzdTh}) is of interest from a mathematical point
of view due to the existence of the quasi--periodic behaviour.  The
orbits are future asymptotic to a {\em heteroclinic cycle}, consisting
of the two saddle equilibrium points $S_1$ and $S_2$ and the single
(boundary) orbits in the invariant sets $z=0$ and $1-x^2-z=0$ joining
$S_1$ and $S_2$ (see Fig. 1). The former set corresponds to the zero
$\Lambda_{\rm M}$ solution (given by Eqs. (\ref{dma}) with $q=0$) and
the latter to the solution with constant axion field (see
Eq. (\ref{invariantset})); to our knowledge this exact solution was
not previously known.  In a given `cycle', an orbit spends a long time
close to $S_1$ and then moves quickly to $S_2$ shadowing the orbit in
the invariant set $z=0$.  It is then again quasi-stationary and
remains close to the equilibrium point $S_2$ before quickly moving
back to $S_1$ shadowing the orbit in the invariant set $1-x^2-z=0$.
We stress that the motion is {\em not} periodic, and on each
successive cycle a given orbit spends more and more time in the
neighbourhood of the equilibrium points $S_1$ and $S_2$.

In Fig. 1, the exact solution corresponding to the equilibrium point $F$ is 
a power-law solution:
\begin{eqnarray}
\label{5.1}
a &=& a_* (-t)^{1/3} \nonumber \\
\Phi &=&\ln \left( \frac{16}{3\Lambda_{\rm M}} \right)-2 \ln (-t) 
\nonumber \\
\sigma &=& \sigma_* \pm \frac{\sqrt{15}\Lambda_{\rm M}}{16} 
(-t)^2 \nonumber \\
\dot{\beta} &=&0  ,
\end{eqnarray}
where $t$ is defined over the range $-\infty < t <0$ by 
a suitable choice of an integration constant. 
This new solution represents a cosmology that collapses monotonically 
to zero volume at $t=0$. The   
curvature and coupling are both singular at this point. 
The universe is initially in a weak coupling regime, since $G_{\rm eff}
\rightarrow 
0$ as $t\rightarrow -\infty$, and the 
effective energy density of the axion field also vanishes in this limit. 

All orbits in Fig. 1 begin at $F$. 
The cyclical nature of these orbits can be physically
understood by reinterpreting the axion field in terms of a 
membrane. The homogeneity of the axion field, 
$\sigma =\sigma (t)$, implies that the two--form potential, 
$B_{\mu\nu}$, must be independent of cosmic time, 
and this in turn implies that its field strength must be proportional 
to the volume form of the three--space. If the topology of the spatial 
sections is given by a three--torus, $S^1 \times S^1 \times S^1$, 
the behaviour of the axion field is dynamically equivalent to that of 
a membrane that has been wrapped around this torus \cite{kaloper}.
The collapse is resisted by this membrane and the universe 
undergoes a bounce. As the volume increases, however, 
the influence of the membrane is diminished, because the energy 
density of the axion is rapidly redshifted away. Consequently, the 
cosmological constant becomes important. 

The subsequent effect of the cosmological constant can be determined
by viewing Eq. (\ref{generalaction}) in terms of a Brans--Dicke
action, where the dilaton--graviton 
coupling parameter is given by 
$\omega =-1$ \cite{bransdicke}.  The Brans--Dicke FRW
models containing only a cosmological constant in the matter sector
have been discussed previously by Barrow and Maeda, but their
solutions only apply for $\omega > -5/6$ \cite{barrow,lidseyprd}.  The
behaviour of the general solution for $\omega < -5/6$ is different and
can be established by performing a conformal transformation to a frame
where the dilaton field is minimally coupled to gravity. In such a
frame the term containing $\Lambda_{\rm M}$ may be viewed as an
exponential, self--interaction potential for the dilaton, where the exponent is
uniquely determined by the value of $\omega$ \cite{ColeyBillyard}.
When $\omega >-5/6$, the late--time attractor is a scaling solution,
where the kinetic and potential energies of the dilaton field redshift
in direct proportion \cite{liddle}.  For $\omega < -5/6$, however, the
potential is so steep that the dilaton effectively becomes massless
\cite{lidseygrg}. Thus, the late--time attractor when $\omega =-1$
corresponds to the solution (\ref{dmvcosmic}) where $h_*^2=1/3$.

Further insight may be gained by defining new variables in 
the reduced action (\ref{taction}):
\begin{eqnarray}
\label{gamma}
\chi \equiv 4\alpha -\Phi \nonumber \\
\gamma \equiv \Phi -6\alpha  \nonumber \\
\tilde{t} = \int dt e^{\varphi}   .
\end{eqnarray}
In the case where $\dot{\beta} =\dot{\sigma} = 0$, 
Eq. (\ref{taction}) reduces to
\be
S=\int d\tilde{t} \left[ -\frac{3}{2} \left( \frac{d\chi}{d \tilde{t}}
\right)^2 +\frac{1}{2} \left( \frac{d\gamma}{d\tilde{t}}
\right)^2  -\Lambda_{\rm M} e^{-\gamma} \right]  .
\ee
The momentum conjugate to the variable $\chi$ 
is constant, i.e., $d\chi /d\tilde{t} =C$, and the field 
equation for $\gamma$ is a Liouville equation: 
\be
\label{gammafield}
\frac{d^2 \gamma}{d \tilde{t}^2} =\Lambda_{\rm M} e^{-\gamma}  .
\ee
The general solution to Eq. (\ref{gammafield}) satisfying 
the Hamiltonian constraint can be found. When 
$C>0$, it can be shown that 
$\gamma \propto  \sqrt{3C^2} \tilde{t}$ in the late--time 
limit. Since the Hubble parameter is given by 
\be
\label{hubble}
\dot{\alpha} =-\frac{1}{2}e^{\varphi} \left( C 
+\frac{d \gamma}{d\tilde{t}} \right) , 
\ee
the late--time attractor corresponds to the {\em collapsing} 
solution in Eq. (\ref{dmvcosmic}). 

In effect, therefore, the  cosmological constant 
resists the expansion and ultimately 
causes the universe to recollapse and asymptotically 
approach the saddle point $S_1$. 
On the other hand, the collapse causes the 
axion field to become relevant once more and a further bounce 
ensues. The 
process is then repeated with the universe undergoing a series of 
bounces. The orbits move progressively 
closer towards the two saddles, $S_{1,2}$, and spend
increasingly more time near to these points. 
This behaviour is related to the fact that the kinetic energy of the shifted
dilaton field increases monotonically with time, since 
Eq. (\ref{r2}) implies that $\ddot{\varphi} > 0$. 

When shear is included $(y \ne 0)$, $F$ still represents the
{\em only} source in the system.  The orbits follow cyclical
trajectories in the neighbourhood of the invariant set $y=0$ and they
spiral outwards monotonically, since Eq. (\ref{yp}) implies that
$dy/d\Theta >0$.  After a finite (but arbitrarily large) number of
cycles the kinetic energy associated with the shear 
parameter, $\beta$, begins to dominate the axion and
cosmological constant. The orbits then asymptote to the
power-law solutions (\ref{dmvcosmic}).  All orbits in
the full three-dimensional phase space actually spiral outwards
around the orbit represented by the dashed line in Fig. 2 which
corresponds to the exact solution (\ref{xfixed})--(\ref{zprime}) with
$x={\rm constant}$.  In terms of cosmic time, $t$, this exact
solution satisfies
\begin{equation}
\Lambda_{\rm M} -\frac{16}{27} \psi^2+\frac{48}{13}N^2=0  \label{B.1}
\end{equation}
and 
\begin{equation}
\dot{\alpha} =-\frac{1}{9}\dot{\varphi}, \label{B.2}
\end{equation}
whence from Eqs. (\ref{r1})--(\ref{frr}) we obtain
\begin{equation}
\label{function}
 \ddot{\varphi}=\dot{\varphi}^2 +k_\varphi^2e^{2\varphi},\label{B.3}
\end{equation}
where $k_\varphi$ is an
integration constant.  Defining 
\begin{equation}
\varrho \equiv \frac{e^{-\varphi}}{k_\varphi} 
\end{equation} 
simplifies Eq. (\ref{function}) to 
\begin{equation}
\label{varrho}
\varrho\ddot{\varrho} = -1
\end{equation}
and Eq. (\ref{varrho}) can be integrated exactly to obtain $\dot \varphi$
\cite{LidseyWaga}. A second integration then yields $\varphi$ in
terms of the Inverse Error function, so that in principle we can
obtain the scale factor as a function of time, $t$, from Eq. (\ref{B.2}).  

This cyclical behaviour does not arise if $\Lambda_{\rm M} < 0$ 
(see Fig. 3). The equilibrium points $A$ and $R$ represent 
the power-law solutions:
\begin{eqnarray}
\nonumber
a & = & \frac{a_*}{\sqrt{\pm\sqrt{-\frac{1}{2}\Lambda_{\rm M}} t}} \\
\nonumber
\Phi & = & \Phi_* -\ln\left[\pm\sqrt{-\frac{1}{2}\Lambda_{\rm M}}t \right]^2  \\
\nonumber
\beta&=&\beta_* \\
\nonumber
\sigma&=&\sigma_* ,\\
\label{at_half}
\end{eqnarray}
where the $+$ sign 
corresponds to the point $R$ and the $-$ sign to 
the point $A$. Initially the universe 
is collapsing and  
the axion field induces a bounce, but this field 
can not dominate the dynamics again once the 
volume of the universe has increased sufficiently. 

Fig. 4 depicts the axisymmetric Bianchi type I model when
$\Lambda_{\rm M} <0$.  In this phase space, for $-1/\sqrt3<u<1/\sqrt3$
the line $W^-$ represents the negative branch of the solution
(\ref{dmvcosmic}) for $h_*\in ( -1/3, 1/3 )$.  Likewise, for
$-1/\sqrt3<u <1/\sqrt3$ the line $W^+$ represents the ``$+$'' solution
in Eq. (\ref{dmvcosmic}) for $h_*\in (- 1/3, 1/3)$. The four saddle
points $S_{u,v}$ correspond to the power-law solutions
(\ref{dmvcosmic}) with $h_*=\pm1/\sqrt 3$. From Fig. 4 we see that
generically trajectories asymptote away from either the line $W^-$ or
the point $R$ and move towards the expanding power-law solutions $W^+$
or $A$. Hence, the cosmological constant is important in determining
both the early-- and late--time dynamics.  Since $u$ is monotonically
increasing (see (\ref{dudXi})) we note that the occurrence of a bounce
in these cosmological models is a typical feature.

\section{Discussion}

In this paper we have presented a qualitative analysis of 
spatially flat FRW and Bianchi type I 
cosmologies containing non--trivial dilaton and 
axion fields with a cosmological constant 
in the matter sector of the theory. The action we considered 
reduces to the string effective action when the cosmological constant 
vanishes. A complete stability analysis 
was performed in all cases by finding variables that 
led to a compactification of the phase space.
We found that 
a cosmological constant has a significant effect on the dynamics 
of the string cosmologies (\ref{dma}). 

One of the more interesting mathematical features of the models
we have considered is the existence of quasi-periodic behaviour.  This
occurs in the isotropic cosmologies, where the 
orbits are future asymptotic to a heteroclinic cycle (see Fig. 1). The
solutions interpolate between the saddles $S_1$ and $S_2$ 
corresponding to the power-law models (\ref{dmvcosmic}) 
with $|h_*| =1/\sqrt{3}$. It would be interesting
to consider the implications of this behaviour for the pre--big bang
inflationary scenario \cite{pbb}.  We note that the phase portrait
depicted in Fig. 1 is similar to that of Fig. 1(e) in \cite{NU}
that describes the locally rotationally symmetric submanifold of the
stationary Bianchi type I perfect fluid models in general relativity,
although in this latter case the independent variable is space-like.

The general Bianchi cosmology, where the shear matrix is given by 
\begin{equation}
\beta_{ab} ={\rm diag} \left[ \beta_+ +\sqrt{3}\beta_- , \beta_+ 
-\sqrt{3} \beta_- , -2\beta_+ \right]
\end{equation}
can be analysed directly by defining the variable $N$ in Eqs. (\ref{4.10}) and
(\ref{y}) via $N^2 = \sum_{i=\pm} N^2_i$.  Orbits in the full phase
space of Fig. 2 with non-trivial shear term
(represented by the variable $y$) are repelled from the source $F$. 
The variable $y$ increases monotonically and the orbits 
spiral around the exact solution given by Eqs. 
(\ref{B.1})--(\ref{B.3}), as represented by the dashed line in
Fig. 2. (See also Fig. 1(f) and the Appendix in \cite{NU}).  This
implies that solutions are asymptotic in the past to the solution
given by Eq. (\ref{5.1}).  At early times the orbits `shadow' the
orbits in the invariant set $y = 0$ and undertake cycles between the
saddles (in three-dimensional phase space) on the equilibrium set $W$
close to $S_1$ and $S_2$. These saddles on $W$ may be interpreted as
Kasner--like solutions \cite{NU,WE}.  Note that $y = 0$ at $S_1$ and
$S_2$, however, and there is no shear term in these
cases.  The orbits thus experience a finite number of cycles in which
the solutions interpolate between different Kasner-like states.  The
orbits eventually asymptote towards a source on the line $W$.

This is perhaps reminiscent of the mixmaster behaviour that occurs
in the Bianchi type VIII and IX cosmologies \cite{WE,hobill}. 
These are the most general models in the 
Bianchi class A of spatially homogeneous universes \cite{em}. 
In these models, Taub
orbits joining equilibrium points of the Kasner set $K$ lead
to the existence of infinite heteroclinic sequences which
approximate the past asymptotic behaviour of generic
orbits. (These heteroclinic sequences are defined by a 
map of $K$ onto itself).  Mixmaster oscillations also
occur in less general (i.e. lower--dimensional) Bianchi
models with a magnetic field \cite{LeBlanc} or
Yang-Mills fields \cite{BL}.  It is interesting to note 
in the string context that mixmaster behaviour also
occurs in scalar-tensor theories of gravity in general
and in Brans-Dicke theory in particular \cite{BDT}.

This analogy is only suggestive.  We note that if a non-zero
central charge deficit is included, the quasi-periodic
behaviour in the full (higher-dimensional) phase space does indeed
persist \cite{Billyard1999}.  Unlike the mixmaster oscillations,
however, the orbits in Fig. 2 eventually spiral away from $y=0$,
although there are orbits that experience a finite but arbitrarily
large number of oscillations.  However, it would be interesting to
further explore any correspondence with possible mixmaster behaviour,
particularly by including addtional anisotropic or matter degrees of
freedom.

Some of the dynamics discussed in this paper is also 
relevant to higher--dimensional cosmological models. 
Kaluza-Klein compactification of ten--dimensional
supergravity theories \cite{gsw} onto an isotropic six-torus
of radius $e^{\beta}$ introduces an additional modulus field 
into the effective four--dimensional action (\ref{sigmaaction}). 
Integration over the spatial variables for a spatially flat FRW model 
then leads to an action 
that is formally identical to that of Eq. (\ref{taction})
when we specify $\Lambda_{\rm M} =0$. In this sense, therefore, 
the action (\ref{taction}) can be
recast into a higher--dimensional context, where 
the shear term $\beta$ plays the role of the modulus field
and $\Lambda_{\rm M}$ is interpreted as a cosmological constant that 
is introduced after compactification. 

More generally, type II supergravity theories 
contain Ramond--Ramond form--fields that do not 
couple directly to the dilaton field in the string frame \cite{gsw}. 
Under dimensional reduction, these fields give rise 
to terms in the effective action of the form $Q^2
\exp(c \beta)$, where $Q$ and $c$ are constants \cite{Scherk1979a};
i.e., Ramond--Ramond charges give rise to exponential potentials for
the modulus field rather than a simple constant term such as 
that considered in this work. However, from the
analysis in section 3.1,  there will be string solutions 
containing Ramond--Ramond fields that 
asymptote towards solutions with $\beta={\rm constant}$
($y=0$ in Fig. 2), in
which case $Q^2 \exp(c \beta)$ is effectively constant. It might then
be expected that the heteroclinic cycle that occurs in the invariant
set $y=0$ (see Fig. 1) will play an important r\^{o}le in describing the
dynamics of these string cosmologies.

\vspace{.3in}

\centerline{\bf Acknowledgments}

\vspace{.3in}
We would like to thank Ulf Nilsson for helpful comments.  
APB is supported by Dalhousie University, AAC is supported
by the Natural Sciences and Engineering Research Council
of Canada (NSERC), and
JEL is supported by the Royal Society.


\end{document}